\documentclass[
               amsfonts, amssymb, amsmath,
	       reprint, showkeys,
	       nofootinbib,
	       prd,
	       floatfix, superscriptaddress, natbib,
	      ]{revtex4-2}

\usepackage[english]{babel}
\usepackage[utf8]{inputenc}
\usepackage{multirow}
\usepackage{float}
\usepackage[pdftex, pdftitle={Article}, pdfauthor={Author}]{hyperref} 
\usepackage{amsthm}
\usepackage{mathtools}
\usepackage{physics}
\usepackage{xcolor}
\usepackage{graphicx}
\usepackage{adjustbox}
\usepackage{placeins}
\usepackage[T1]{fontenc}
\usepackage{lipsum}
\usepackage{csquotes}
\usepackage{sidecap}
\usepackage[normalem]{ulem}
\usepackage{enumitem}
\usepackage{relsize}

\newcommand{\aap}{Astron. Astrophys.}
\newcommand{\apjl}{Astrophys. J. Lett.}
\newcommand{\apjs}{Astrophys. J. Suppl.}
\newcommand{\jcap}{J. Cosmol. Astropart. Phys.}
\newcommand{\mnras}{Mon. Not. Roy. Astron. Soc.}

\newcommand{\pasj}{Publ. Astron. Soc. Jpn.}

\newcommand\Mpc{h^{-1}\text{Mpc}}
\newcommand\Mpcinv{h\text{Mpc}^{-1}}

\newcommand\Rvir{R_\text{vir}}
\newcommand\Mvir{M_\text{vir}}
\newcommand\Msun{M_\odot}

\begin{document}

\title{De-baryonifying halos via optimal transport}

\author{Leander Thiele}
\email{leander.thiele@ipmu.jp}
\affiliation{Center for Data-Driven Discovery, Kavli IPMU (WPI), UTIAS, The University of Tokyo, Kashiwa, Chiba 277-8583, Japan}
\affiliation{Kavli IPMU (WPI), UTIAS, The University of Tokyo, 5-1-5 Kashiwanoha, Kashiwa, Chiba 277-8583, Japan}

\begin{abstract}
  Baryonic feedback uncertainty is a limiting systematic for next-generation weak gravitational lensing analyses.
  At the same time, high-resolution weak lensing maps are best analyzed at the field-level.
  Thus, robustly accounting for the baryonic effects in the projected matter density field is required.
  Ideally, constraints on feedback strength from astrophysical probes should be folded into
  the weak lensing field-level likelihood.
  We propose a macroscopic method based on an empirical correlation between feedback strength
  and an optimal transport cost.
  Since feedback is local re-distribution of matter, optimal transport is a promising concept.
  In this proof-of-concept, we de-baryonify projected mass around individual halos in the IllustrisTNG simulation.
  We choose the de-baryonified solution as the point of maximum likelihood on the hypersurface
  defined by fixed optimal transport cost around the observed full-physics halos.
  The likelihood is approximated through a normalizing flow trained on multiple gravity-only simulations.
  We find that the set of de-baryonified halos reproduces the correct convergence power spectrum suppression.
  There is considerable scatter when considering individual halos.
  We outline how the optimal transport de-baryonification concept can be generalized to full convergence maps.
\end{abstract}

\maketitle

\section{Introduction}

Baryonic feedback is a major unknown in the interpretation of the late-time matter distribution.
Weak gravitational lensing is particularly affected by this systematic:
the suppression of small-scale power due to baryonic feedback can easily be confused with a
lower amplitude of clustering in the initial conditions.
This partial degeneracy may be responsible for the apparent mismatch in measurements of the clustering
amplitude $S_8$ between the cosmic microwave background and weak gravitational lensing~\citep[e.g.,][]{Amon2022a,Hadzhiyska2024,Elbers2024,Bigwood2024},
although there are hints to the contrary as well~\citep[e.g.,][]{Grandon2024,Terasawa2024,Salcido2024}.
Relatedly, accurate accounting for baryonic feedback will be necessary for trustworthy constraints
on neutrino mass from weak lensing~\citep[e.g.,][]{Fong2019,Liu2019,Ferlito2023}.

Theoretically, baryonic feedback is poorly understood.
The reason is the separation of scales: gravitational energy is converted into thermal and kinetic
energy on astrophysical scales.
The primary mechanism are probably the super-massive black holes residing at the centers of most massive
galaxies.
This small-scale energy is then transferred outwards and partly converted back into gravitational energy
through the displacement of mass.
Describing this entire process with bottom-up techniques is very challenging and even high-resolution
hydrodynamic cosmological simulations need to adopt an effective coarse-grained description.

Therefore, analyses of weak lensing data resort to effective methods.
The existing techniques can be divided into three classes.
First, scale cuts~\citep[e.g.,][]{Asgari2021,Amon2022,Dalal2023,Li2023}.
The most conservative technique identifies below which length scale the gravity-only approximation
can no longer be trusted (relying on hydrodynamical simulations).
Since baryonic feedback and cosmological information are not completely degenerate, this method necessarily
loses constraining power from the small scales.
Second, descriptions at the level of summary statistics~\citep[e.g.,][]{Mead2015,Mead2016,Huang2019,Mead2021,Huang2021}.
Typically, these only attempt to describe the feedback effects on the two-point function.
Particularly noteworthy is the empirical (but physically expected) relationship between the power spectrum
suppression and the amount of retained gas~\citep{vanDaalen2020,vanLoon2024}.
Such relations are unable to make predictions about higher order statistics and therefore render analyses sub-optimal.
Third, field-level methods.
These are necessary in order to enable optimal field-level inference~\citep[e.g.,][]{Porqueres2023,Dai2024}.
The primary example is the baryonification concept~\citep{Schneider2015,Schneider2020,Arico2020,Arico2024,Anbajagane2024}. 
This technique displaces particles in gravity-only simulations to match halo density profiles that include
baryonic feedback.
It has been shown that baryonification can reproduce the two- and three-point functions from various
hydrodynamic simulations at the expense of introducing a handful of nuisance parameters~\citep{Arico2021}.
Higher order statistics such as peak counts may present a challenge to current baryonification parameterizations~\citep{Lee2023}.
Analogously to the halo occupation distribution, dependences beyond halo mass and stochasticity
could be necessary in order to fulfill the accuracy requirements of Stage-IV surveys.
Other field-level methods utilize the enthalpy gradient~\citep{Dai2018} or the transfer function~\citep{Sharma2024}.

Aside from the conservative scale cuts approach, it appears that baryonification is likely to become the
default method to deal with baryonic feedback in Stage-IV weak lensing analyses~\citep{Laureijs2011,Takada2014,Spergel2015,Ivezic2019}.
Baryonification has clear advantages.
It operates at the field level, thus not restricting the optimality of cosmological constraints in principle.
It has a clear physical interpretation; this simplifies the construction of priors for its nuisance parameters
and enables making contact with astrophysical probes of baryonic feedback.
It does not require training on hydrodynamic simulations.

In this work, we take a step back and attempt to formulate a concept for baryonic feedback that
borrows from the strengths of baryonification but approaches the problem from the opposite direction in the
following sense.
Baryonification (and other existing approaches) start from the gravity-only prediction and iteratively add
modifications through the introduction of nuisance parameters.
Once all existing hydrodynamic simulations can be reproduced at some desired level of accuracy and for a
given set of summary statistics, the model is declared sufficient.
Thus, ignorance is gradually added to the model.
We propose to start with maximum ignorance and gradually add knowledge.

What do we know about baryonic feedback?
It originates in halos, where energy is injected into the matter distribution by radially displacing particles outwards.
How much energy is injected, and what level of isotropy this displacement has, we do not know very well.
Baryonic feedback conserves mass to a good approximation, thus inspiring the following:
We propose to make a connection between energy input and optimal transport cost.
An optimal transport plan is a set of matter displacements that minimizes a cost function to be defined
and transforms the full-physics matter field into its hypothetical gravity-only counterpart.
Since weak lensing only probes projected density, the correspondence between optimal transport cost
and energy injection is necessarily stochastic.
Furthermore, optimal transport cost is a real metric whereas energy difference can have both signs.
Thus, we cannot give a robust theoretical framework for the connection, but will attempt to provide
evidence that it is useful.

Optimal transport has been used with similar motivation in galaxy clustering~\citep{Nikakhtar2022,Nikakhtar2023,Nikakhtar2024}.

\begin{figure}
  \includegraphics[width=\linewidth]{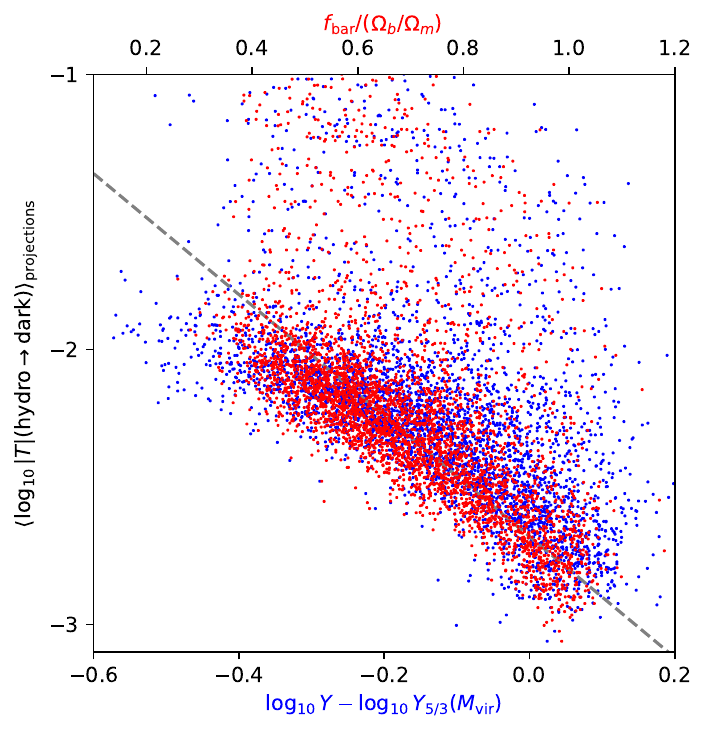}
  \caption{
    Illustration of the relationship between optimal transport cost and feedback strength.
    The twin horizontal axes display halo properties that are known to be anti-correlated with feedback.
    In blue, the deviation of logarithmic integrated electron pressure within $\Rvir$
    from the self-similar $M_\text{vir}^{5/3}$ relationship.
    In red, the baryonic mass fraction within $\Rvir$ relative to the cosmic mean.
    The vertical axis displays the logarithmic optimal transport cost between the full-physics and gravity-only
    matter (as defined in the text), averaged over three projections.
    Each data point is an individual halo in the Illustris TNG 300 simulation;
    the dashed line is for illustration purposes only.
    The optimal transport cost is correlated with energy input.
    One might wonder about halo mass as a confounding factor; at fixed $\Delta Y$ no correlation
    between halo mass and optimal transport cost is visible.
  }
  \label{fig:costy}
\end{figure}

As empirical support for the connection between baryonic feedback and optimal transport,
Fig.~\ref{fig:costy} shows the optimal transport cost between full-physics and gravity-only halos in a simulation,
plotted against two proxies for baryonic feedback strength.
In blue, the deviation of the thermal Sunyaev-Zel'dovich $Y$-parameter from the self-similar $\Mvir^{5/3}$ relation.
In red, the baryon fraction relative to the cosmic mean.
Both $\Delta Y$~\citep{Wadekar2023,Pandey2023} and $f_\text{bar}$~\citep{vanDaalen2020,MartinAlvarez2024} are known to be anti-correlated with baryonic feedback strength.
Thus, Fig.~\ref{fig:costy} demonstrates that the optimal transport cost is indeed a (noisy) proxy for energy input.
We will give the full definition of the transport problem's formulation later.

One point of uncertainty is the ratio between thermal and kinematic feedback modes.
We expect that this ratio will impact the relationship between total energy input and optimal transport cost.
We assume that multi-wavelength observations will be able to disentangle between the feedback modes.
One promising combination of measurements could be $f_\text{gas}$ and tSZ $Y$.

How can we make the described connection between optimal transport cost and feedback strength practically usable?
We cannot give the complete answer yet.
A qualitative plan would be to take a measured weak lensing mass map and an estimate of the baryonic energy input
(this estimate could be global or have some local modulation).
Using results from hydrodynamical simulations, we could translate the energy input into an optimal transport cost.
Equipped with such a cost estimate, we would find the gravity-only mass map which has the required optimal transport
distance from the observed map while maximizing the probability in the distribution of mass maps.
This distribution could be conditioned on cosmological parameters, enabling a full parameter inference pipeline.

This qualitative picture is far from complete and will take more work to flesh out.
In this work, we restrict ourselves to a related but easier problem: instead of full convergence maps,
we debaryonify individual halos.
Furthermore, we restrict to redshift $z=0$ and fixed cosmological model.
Due to the fixed cosmology, at the level of statistical averages our problem is trivial to solve.
However, the field-level setup makes it non-trivial.
The precise setup will be described in the next section.

De-baryonification has the advantage that it does not require us to assume a specific parameterization of
the effects of baryonic feedback.
Indeed, it is clear that any ad-hoc model will not cover the exact realization of baryonic feedback implemented
in our universe.
However, we may hope that in the high-dimensional space of mass maps the posterior over gravity-only maps
compatible with the observation and a given level of energy input will be extremely sharply peaked.

\section{Methods}

We write the likelihood over gravity-only mass maps $x_d$ given an observed full-physics map $x_h$ schematically as
\begin{equation}
  p(x_d | x_h) \propto p(x_d) p(|T_{hd}|)\,,
  \label{eq:likeschema}
\end{equation}
where $p(x_d)$ is the distribution of gravity-only maps, and $T_{hd}$ is the optimal transport plan from the
full-physics to the gravity-only map.
The norm $|T_{hd}|$ is the optimal transport cost whose likelihood $p(|T_{hd}|)$ we will describe later.
Eq.~\eqref{eq:likeschema} is only schematic and motivates the following description;
the actual likelihood is given in Eq.~\eqref{eq:like}.

Analytically, the field-level likelihood $p(x_d)$ is difficult to obtain since it contains highly non-linear physics.
Thus, we use a deep generative model to learn this distribution from simulations.
Note that these simulations are gravity-only, thus relatively cheap and, crucially, free of modeling uncertainty.

If $x_d$ were a full weak lensing mass map, learning $p(x_d)$ would be a tractable but non-trivial challenge.
The complications include limited availability of training data, and the translational and rotational symmetries.
Since the focus of this work is more on the optimal transport plan of the likelihood, we simplify the deep learning
problem by replacing $x_d$ with mass maps of individual halos.
Besides drastically simplifying the construction of $p(x_d)$, this shift in problem specification also makes the
initial exploration presented in this work more interpretable.

\subsection{Data}
\label{sec:data}

We define $x_d$ as the two-dimensional projection of the matter field around halos within $3\,\Rvir$.\footnote{
  We extract the data from simulations using the code at \url{https://github.com/leanderthiele/group_particles}.
}
The radial cutoff is motivated by the fact that we use the TNG subgrid model as the test set;
the baryonic feedback in TNG does not typically displace mass beyond $3\,\Rvir$.

Halos are chosen as Rockstar-identified~\citep{Behroozi2013} host halos with masses $\Mvir > 10^{13}\,h^{-1}\Msun$.
The mass cutoff is chosen because it keeps computational and storage costs manageable and at this threshold
the TNG model already displays pronounced baryonic feedback effects.
Furthermore, it is known that group-size halos are the primary sources of power spectrum suppression at the
relevant scales of $k \sim 1\,\Mpcinv$~\citep{vanLoon2024}.
When working with the full-physics halos in the IllustrisTNG simulation,
we use the coordinates and radii computed by Rockstar in the gravity-only simulation.

Even though the TNG model does not appreciably affect the matter distribution beyond $3\,\Rvir$,
mass is not exactly conserved (the sums of $x_d$ and $x_h$ between paired halos are not exactly the same).
The level of mass non-conservation is relatively minor (percent level, typically).
However, it would still make the optimal transport problem more difficult to formulate.
Therefore, we normalize the maps $x$ to unit mass.

We choose to represent $x$ in polar coordinates.
We bin the mass into 16 radial and 16 angular bins, centered on the halo center computed by Rockstar.
The angular bins are naturally equally spaced, while the radial bin spacing is chosen such that each
bin contains approximately equal mass on average.

\subsection{Normalizing flow}

As described above, we need to learn the distribution $p(x_d)$ with a deep generative model.
For this task, we choose the normalizing flow architecture.
It provides an explicit, unbiased, efficient density and is thus more suitable to the problem
than other generative setups.
Since the density estimator's quality is essential for the results of this work,
we will present training and performance in sufficient detail.

\subsubsection{Data transformation}

Due to the normalization to unit mass, the data dimensionality is $16\times16 - 1 = 255$.
Since all elements of $x_d$ are physically expected to be non-zero positive, $x_d$ takes values on
the standard simplex.\footnote{In practice, individual pixels can be empty, particularly at low
resolution. In this case, we fill them in with a small positive number.}
It is beneficial to map the $256$ dimensional simplex to $\mathbb{R}^{255}$.
We perform this mapping using the isometric log-ratio (ILR) transform~\citep{Aitchison1982}.
First, each $x_i$ is mapped to
\begin{equation}
  x_i' = \log(x_i) - \frac{1}{256}\sum_{j=1}^{256}\log(x_j)\,.
\end{equation}
Note that this transformation is invertible using the softmax.
Then, the $x_i'$ are projected onto $255$ orthonormal basis vectors $\mathbf{e}_k$
that span the sub-space orthogonal to $\{1\ldots1\}_{256}$.
We compute the basis vectors using an SVD on a set of simulated samples.
The resulting transformed $x$ have a much more Gaussian distribution.
This simplifies the deep learning problem.

\subsubsection{Training data}

Constructing the training set is slightly non-trivial.
We will use the Illustris TNG-300 simulation~\citep{Nelson2018,Marinacci2018,Naiman2018,Pillepich2018,Springel2018,Nelson2019} as a test set.
Therefore, ideally we need to match both resolution and cosmological parameters in our training set.
Among the publically available gravity-only simulations, we find miniUchuu~\citep{Prada2023,Aung2023,Dong-Paez2024,Oogi2023,Ishiyama2021} to be the ideal choice
as its cosmology matches Illustris TNG and its resolution is very close to the medium-resolution (1250)
run of TNG.
However, due to the relatively small box size of $400\,\Mpc$, miniUchuu is not sufficient.
For most generative tasks with the described data it would probably be, but we require our normalizing flow
to have excellent out-of-distribution performance as well as smooth gradients (for efficient sampling).

Therefore, we pre-train the model with a range of other simulations.
These are the high-resolution fiducial boxes of Quijote~\citep{Villaescusa-Navarro2020}, MDPL2 and SMDPL~\citep{Prada2012}, and Uchuu.
For Uchuu, only $0.5\,\%$ of the particle catalog were saved, making it similar in resolution to Quijote.
We include samples from all these simulations in roughly equal proportions during pre-training.
We find that including this pre-training improves validation performance marginally
but yields more cleanly structured latent spaces.

Prior to training, we normalize the samples to zero mean and identity covariance.
For each simulation, the normalization is performed separately.
The exception is, of course, the Illustris test set for which we apply the same normalization
as for miniUchuu (our final fine-tuning training set).

We resample the halo mass function to balance the training set.
This is naturally possible by picking more random projections of higher-mass halos.
The training set has constant $dn/d\log\Mvir$ up to $10^{14.5}\,h^{-1}\Msun$.

Equipped with this training set, we learn the joint distribution $p(x_d, \Mvir)$
using a masked auto-regressive flow~\citep{Papamakarios2017,Huang2018}.\footnote{
  For the MAF, we use the implementation in \href{https://zuko.readthedocs.io}{\texttt{zuko}}.
}
As expected, we find a deep model with approximately $400M$ parameters to yield the best validation loss.

\subsubsection{Inspecting the trained model}

Sampling from the trained normalizing flow yields the correct mean and covariance matrix.

\begin{figure}
  \includegraphics[width=\linewidth]{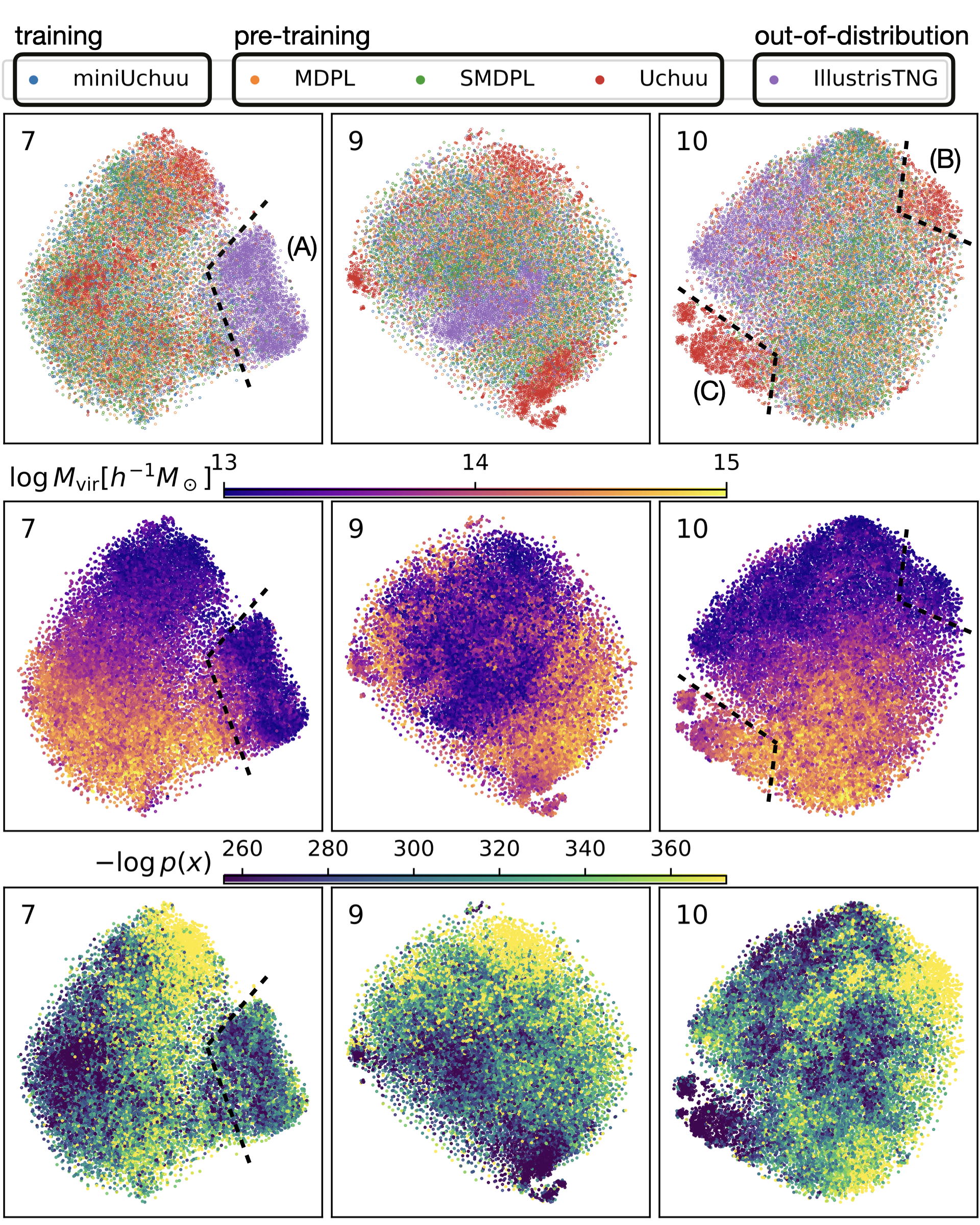}
  \caption{
    Visualizations of latent space in our generative model for gravity-only halos.
    The compression from 256 to 2 dimensions is performed using tSNE.
    From left to right, we progress through the normalizing flow.
    The right-most column corresponds to the base-density.
    From top to bottom, points are color-coded by simulation, halo mass, and inferred probability density.
    The regions A, B, C indicated in the embeddings are noteworthy features further discussed in the main text.
    Note that we randomized the plotting order, so there is no ``overlap bias''.
  }
  \label{fig:latent}
\end{figure}

In Fig.~\ref{fig:latent}, we show visualizations of the normalizing flow latent space
when it is evaluated on samples from the held-out validation sets of various simulations.
At three different points within the flow, we embed the activations in a two-dimesional space using tSNE~\citep{vanderMaaten2008}.
Note that due to the flow's bijectivity, theoretically there exists a dimensionality reduction algorithm
that yields the same embedding at all points in the flow.
However, this is not true for tSNE.
Therefore, we pick three visualizations that most clearly bring out the data structure.

Focusing on the region marked (A), we observe that the out-of-distribution full-physics halos from the
IllustrisTNG simulation indeed fall into a separate part of latent space.
The connection between the TNG part and the rest of latent space is at the high-mass end.
Since baryonic feedback is less pronounced for higher-mass halos due to the deeper graviational potential wells,
this structure is physically expected.
Furthermore, since each sample extends out to $3\,\Rvir$, for higher-mass halos the small-scale details are less important.
Note, however, that the probability density assigned to the TNG halos is within distribution (bottom panel).
This failure to penalize out-of-distribution samples is a known weakness of normalizing flows~\citep{Nalisnick2019}.
Energy-based models might be a better choice in this respect~\citep{Du2019}, but they are usually difficult to train.
When sampling from the flow, as described later, we initialize the chains in the gravity-only sector
and do not observe indications that the poor out-of-distribution performance impacts our results.

In region (B), we see that at the low-mass end the lower resolution of both Uchuu and MDPL makes
the corresponding halos outliers.
Interestingly, in this case the flow is able to penalize the out-of-distribution samples.
This may be a consequence of the pre-training.

Region (C) shows the medium-mass part of Uchuu.
Since Uchuu has the lowest resolution amongst the simulations included in the plot,
it makes sense that even at medium halo masses the samples are out-of-distribution.
However, the assigned probability densities are high, similar to region (A).
As expected, the transition between region (C) and the bulk of latent space
is at the high-mass end where the resolution effect is less important.

In summary, the latent space exhibits rich structure with clear physical interpretation.
The poor out-of-distribution performance of the normalizing flow does not appear to impact our results
but should be kept in mind for similar applications.

\subsection{Optimal transport}

The classical optimal transport problem consists in finding a matrix $T_{hd} \in \mathbb{R}^{256\times256}$
such that its row(column)-wise sums equal $x_h$ and $x_d$, respectively, and the transport cost
\begin{equation}
  |T_{hd}| \equiv \tr M^T T_{hd}
\end{equation}
is minimized for a given choice of cost matrix $M$.
The cost matrix element $M_{ab}$ equals the cost of transporting unit mass from point $a$
in the input to point $b$ in the output.

Efficient algorithms exist for this problem.
However, they are not amenable to parallel execution benefitting from GPU acceleration.
Since we need to solve the problem many times for many different halos, we choose to modify it by adding
an entropy regularization term.
Including this term makes it possible to apply the Sinkhorn algorithm~\citep{Sinkhorn1967}, which can easily be batch-evaluated
on GPUs.
We utilize the recently-developed simulated annealing-like version of the Sinkhorn algorithm
with efficient gradients~\citep{Feydy2019}.

For the cost matrix, we use the squared Euclidean metric
\begin{equation}
  M_{ab} = \frac{1}{2} ||\mathbf{r}_a - \mathbf{r}_b||^2\,,
\end{equation}
where $\mathbf{r}$ are two-dimensional vectors.
Our results are somewhat sensitive to the choice of cost matrix.
In particular, the square-rooted version leads to an incorrect prediction of the power spectrum suppression.
Thus, the presented methodology does indeed perform something non-trivial.
On the flip-side, it is not obvious why the cost matrix should generalize to different implementations
of baryonic feedback (compared to our Illustris-TNG test set).
We leave this question to future work.

We choose the regularization term relatively small and find the resulting optimal transport cost to be
very close to the exact zero-entropy solution.

\subsection{Likelihood}

Using the trained density estimator $p(x_d, \Mvir)$ and the optimal transport implementation for $|T_{hd}|$,
we write the likelihood as
\begin{equation}
  p(x_d | x_h, \Mvir, \mathcal{T}) \propto
    p(x_d | \Mvir)
    e^{-\frac{1}{2} \left( \frac{ \log(|T_{hd}|/\mathcal{T}) }{\sigma_{\log T}} \right)^2}
    |T_{hd}|^{-\alpha}\,,
  \label{eq:like}
\end{equation}
where $\mathcal{T}$ is the transport cost evaluated on the true $x_d$ from the simulation,
and $\sigma_{\log T} = 0.01\,\text{dex}$.
The power law term is added to counteract the volume effect:
with increasing $|T_{hd}|$, there are more transport plans available.
We use $\alpha = 10$, but our results are not very sensitive to this exact choice.
This likelihood approximates taking a slice of fixed $|T_{hd}|=\mathcal{T}$ through the distribution $p(x_d | \Mvir)$.
We were unable to find a construction that computes this slice exactly, but with a spread of only $0.01\,\text{dex}$
the approximation is very good.

Directly maximizing this likelihood can yield pathological out-of-distribution results.
The reason is that the minimizer is able to find very narrow spots in latent space where
the normalizing flow returns abnormally high density.
Monte Carlo sampling the likelihood, however, stays within the distribution and gives as
an added bonus the full posterior instead of just the peak.

Therefore, we use Hamiltonian Monte Carlo sampling on the above likelihood.
We sample in the ILR transformed space.
By adapting the mass matrix and tweaking the leapfrog step size and length, we obtain
an acceptance rate of typically $65\,\%$ with auto-correlation time of $\sim 200$.
For each halo, we obtain $16384$ samples and discard the first $4000$ as a conservative burn-in.
Then, we use the sample with the maximum likelihood as the MAP point.
The obtained MAP points do not show the same pathological OOD characteristics as were obtained with
direct optimization.

\section{Results \& Discussion}

As described in the previous section, equipped with the density estimator $p(x_d, \Mvir)$ and the optimal transport algorithm,
we sample the likelihood Eq.~\eqref{eq:like} for each of the $3 \times 3926$ test halos in the Illustris TNG-300 simulation at redshift $z=0$.
There are three versions per halo, corresponding to the three coordinate axes along which we project the matter field.
We take the highest-resolution full-physics simulation as input $x_h$, but the mid-resolution gravity-only simulation
as the ground truth $x_d$ as it matches the miniUchuu training set better.

\begin{figure}
  \includegraphics[width=\linewidth]{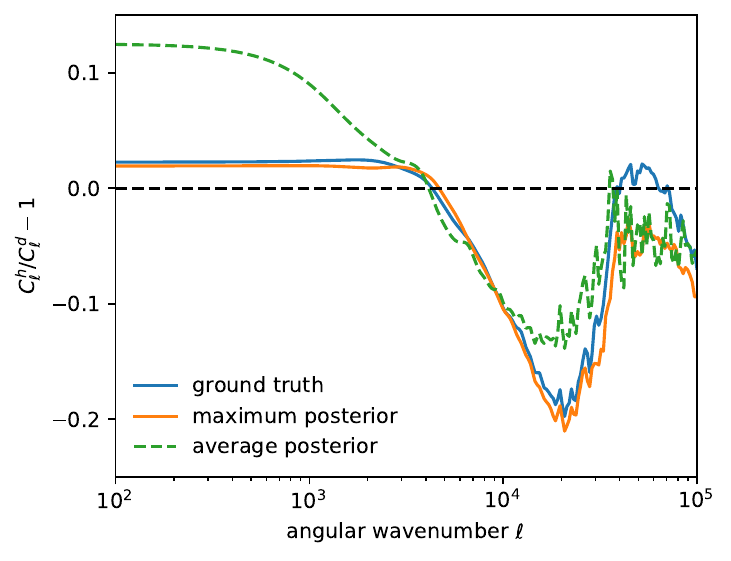}
  \caption{
    The power spectrum suppression computed as described in the text.
    The maximum posterior samples reproduce the true suppression very well,
    while the average does not.
    Note how the ``noise'' matches between the lines, indicating that we
    have successfully constructed a field-level model instead of just
    reproducing the mean of the summary statistic.
  }
  \label{fig:onehalo}
\end{figure}

Using the obtained samples, we compute the suppression in the angular power spectrum.
Since we do not have real weak lensing maps, this is done as follows.
We pretend our halos to be at redshift $z=0.5$ (by keeping the physical size identical).
Then, we compute the ``one-halo term'' for each halo using the standard expression~\citep{Cooray2000}.
We integrate out to $\Rvir$ using the composite Simpson rule.
Note that since the projection goes out to $3\,\Rvir$ the calculation is not exactly equivalent to the usual one-halo prescription.
Our results are qualitatively insensitive to the exact choice of integration boundary.
At this point we need to invert the normalization to unit mass.
Our results are rather insensitive to whether we use the gravity-only or full-physics normalization constant,
due to high level of mass conservation within $3\,\Rvir$.
Then we simply add all one-halo terms up, obtaining a Monte Carlo estimate of the mass integral
(bounded from below by our cutoff of $\Mvir > 10^{13}\,h^{-1}\Msun$).

The result of this approximate computation is shown in Fig.~\ref{fig:onehalo}.
We see that the maximum posterior samples give a very good approximation of the true power spectrum suppression.
On the other hand, the average over samples does not.
In particular, the average does not even reproduce the large-scale power spectrum correctly.
This indicates strong contamination by large volumes with low probability configurations
(a ``prior volume effect'').

\begin{figure}
  \includegraphics[width=\linewidth]{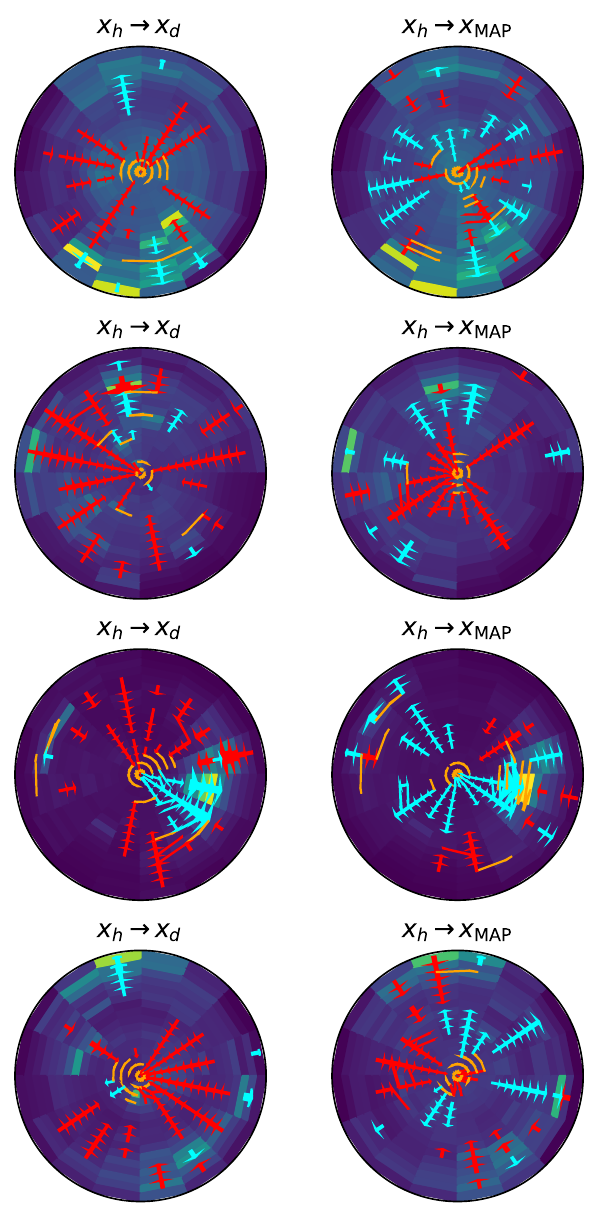}
  \caption{
    Some randomly chosen example halos. Left column shows the transport plan to the true simulated gravity-only halo,
    while the right column shows the plan to the MAP sample from our chains.
    Radially inflowing matter is shown with red arrows, outflowing with cyan, and rotating with orange.
    For clarity, we only show the dominant matter transfers.
    The radial scale is the one adopted in our radial binning, so does not directly correspond to physical length.
    The outer edge in the plots is at $3\,\Rvir$, while $1\,\Rvir$ is approximately half-way.
  }
  \label{fig:plans}
\end{figure}

At the level of summary statistics, such as the convergence power spectrum (Fig.~\ref{fig:onehalo}),
the debaryonification concept appears to work well.
How does it perform when considering individual halos?
In Fig.~\ref{fig:plans}, we show some randomly chosen example halos.
The background color is the projected mass within $3\,\Rvir$, with the special radial binning as described
in Sec.~\ref{sec:data}.
In the foreground, we illustrate the optimal transport plans from full-physics to gravity-only using arrows.
Red arrows point inwards (i.e., in the ``expected'' direction), while blue arrows point outwards.
The left column shows the true plans obtained between matched simulated halos,
while the right column shows the plans obtained with the de-baryonified maximum posterior halos.
First, we observe relatively little dipolar structure, indicating that the shift in center of mass is not
very relevant.
Second, there are substantial differences between the ``true'' and de-baryonified halos.
This indicates that at the level of individual halos there are multiple plausible gravity-only configurations
consistent with a fixed optimal transport cost.

\begin{figure}
  \includegraphics[width=\linewidth]{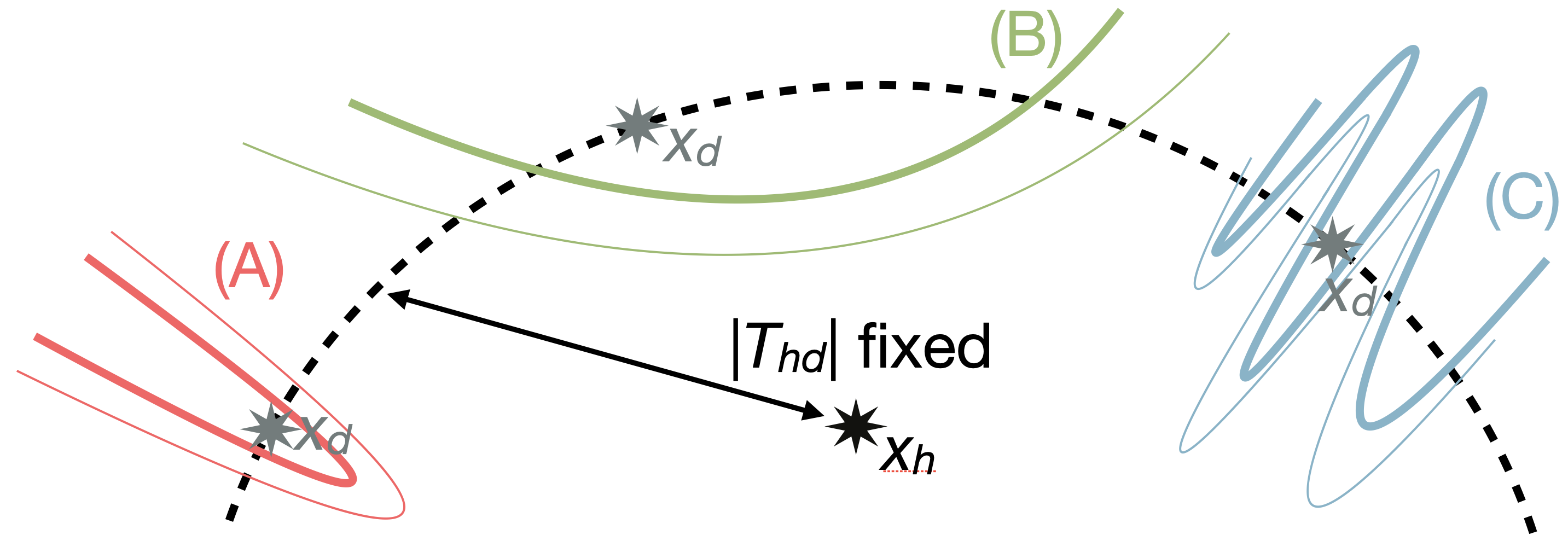}
  \caption{
    Illustration of the structure of probability space.
    Three cases A-C illustrate different possibilities how a slice of constant optimal transport cost,
    originating from a sample $x_h$, can be structured in probability space.
  }
  \label{fig:sketch}
\end{figure}

Schematically, Fig.~\ref{fig:sketch} depicts different plausible scenarios for the structure of probability space.
The point $x_h$ is the full-physics halo, around which the dashed line depicts the codimension-one hypersurface
of optimal transport cost.
The posteriors labeled A, B, C, show different structures of probability space.
We believe scenario (C) to be most accurate, with multiple pockets of high probability intersecting with the
fixed transport cost slice.
It is quite plausible that the generalization to weak lensing mass maps will have similar structure,
indicating a need to go beyond the maximum-posterior solution used in this work.
Note, however, that high-dimensional geometry often defies intuition, thus Fig.~\ref{fig:sketch} should be
taken with a grain of salt.

\section{Conclusions}

The aim of this work is to develop a macroscopic field-level description of baryonic feedback
in order to perform robust and constraining field-level weak gravitational lensing analyses.
By ``macroscopic'', we mean a direct correspondence between astrophysical measures of feedback
strength (such as tSZ and halo gas fraction) and the displacement of mass relevant for weak lensing.
We find that there exists an empirical correlation between feedback strength and an optimal transport cost.
Motivated by this finding, we develop a simple proof-of-concept in which we translate full-physics
halos to their hypothetical gravity-only counterparts by considering the induced probability
on the hypersurface of fixed optimal transport cost around the full-physics oberved halo.

We find that at the level of averages, namely the convergence power spectrum, this de-baryonification
works well and recovers the correct power spectrum suppression.
This is a non-trivial result, despite being restricted to fixed cosmology thus far.

Macroscopic de-baryonification operates naturally at the field level.
Furthermore, it adopts a stance of maximum ignorance, enabling gradual additions of prior knowledge.
This philosophy contrasts with the microscopic concept of baryonification, where one must hope
that sufficient small-scale parameters will cover all macroscopic outcomes.
Additionally, connecting the small-scale parameters of baryonification with astrohpysical
feedback measures is a very challenging problem and requires high-fidelity hydrodynamical
simulations.

Our approach is in its infancy and requires much additional work.
In particular, the relationship between optimal transport cost and feedback strength
needs to be investigated across different feedback implementations.
If it is too dependent on subgrid models, the macroscopic advantage would be lost.

There is no obvious lower-dimensional approximation to the presented de-baryonification.
It always requires a full field-level likelihood and is difficult to interpret.
One possible simplification could be replacing the optimal transport evaluation with
a diffusion model~\citep[e.g.,][]{tong2021diffusion,albergo2023stochastic}, potentially allowing for more fine-grained control.

Finally, it is unclear how to tighten the prior on our understanding of feedback.
Baryonification contains some prior on halo mass dependence, which follows from robust physics.
It is not clear how such a prior (and thus the corresponding coupling with the cosmological lensing
signal) could be incorporated in de-baryonification.

An interesting avenue to explore would be de-baryonification without an external prior on the optimal transport cost.
In principle, this should be possible with a Bayesian approach.
However, it does require computing a ``volume term'', denoting the differential element
in configuration space corresponding to an increment in transport cost.
A power-law approximation has been used in Eq.~\eqref{eq:like}.
We have been unable to perform this computation but hope to take it up in future work.

In summary, de-baryonification via optimal transport shows some interesting promise in a proof-of-concept.
We hope it will be possible to refine it to a true macroscopic description of feedback.

As a by-product, we find that normalizing flows can benefit from extensive pre-training.
Out-of-distribution performance is poor, however;
applications of normalizing flows in science need to anticipate such deficiencies.
We recommend using visualizations (e.g., using tSNE) of latent activations to gain intuition
about a generative model's structure.

\acknowledgments
I thank Farnik Nikakhtar for the discussion that inspired this work.
I thank Jean Feydy and Bruno L\'evy for helping with questions about optimal transport and the Sinkhorn solver.
I thank Adrian Bayer, Carolina Cuesta-Lazaro, Eiichiro Komatsu, Jia Liu, Fabian Schmidt, David Spergel, and Masahiro Takada for useful discussions.
I thank Masahiro Takada for comments on an earlier draft.

This research used computing resources at Kavli IPMU.
The Kavli IPMU is supported by the WPI (World Premier International Research Center) Initiative of the MEXT (Japanese Ministry of Education, Culture, Sports, Science and Technology).
This work was supported by JSPS KAKENHI Grant 24K22878.

We thank Instituto de Astrofisica de Andalucia (IAA-CSIC), Centro de Supercomputacion de Galicia (CESGA) and the Spanish academic and research network (RedIRIS) in Spain for hosting Uchuu DR1, DR2 and DR3 in the Skies \& Universes site for cosmological simulations. The Uchuu simulations were carried out on Aterui II supercomputer at Center for Computational Astrophysics, CfCA, of National Astronomical Observatory of Japan, and the K computer at the RIKEN Advanced Institute for Computational Science. The Uchuu Data Releases efforts have made use of the skun@IAA\_RedIRIS and skun6@IAA computer facilities managed by the IAA-CSIC in Spain (MICINN EU-Feder grant EQC2018-004366-P).

The CosmoSim database used in this paper is a service by the Leibniz-Institute for Astrophysics Potsdam (AIP). The MultiDark database was developed in cooperation with the Spanish MultiDark Consolider Project CSD2009-00064.
The authors gratefully acknowledge the Gauss Centre for Supercomputing e.V. (www.gauss-centre.eu) and the Partnership for Advanced Supercomputing in Europe (PRACE, www.prace-ri.eu) for funding the MultiDark simulation project by providing computing time on the GCS Supercomputer SuperMUC at Leibniz Supercomputing Centre (LRZ, www.lrz.de). 

\bibliography{main}

\end{document}